\documentclass[a4paper]{amsart}

\usepackage{amssymb,amsmath,graphics,epsfig}
\usepackage{latexsym}
\usepackage{eucal}
\makeatletter
\@addtoreset{equation}{section}\makeatother

\newtheorem{theo}{Theorem}[section]
\newtheorem{lem}[theo]{Lemma}
\newtheorem{prop}[theo]{Proposition}
\newtheorem{cor}[theo]{Corollary}

\newcommand{\norm}[1]{\left\Vert#1\right\Vert}
\newcommand{\beq}{\begin{equation}}
\newcommand{\eeq}{\end{equation}}
\newcommand{\bce}{\begin{center}}
\newcommand{\ece}{\end{center}}
\newcommand{\barr}{\begin{array}}
\newcommand{\earr}{\end{array}}
\newcommand{\ben}{\begin{enumerate}}
\newcommand{\een}{\end{enumerate}}
\newcommand{\li}{\mathcal{L}}
\newcommand{\rr}{\mathbb{R}}
\newcommand{\nn}{\mathbb{N}}
\newcommand{\cc}{\mathbb{C}}
\newcommand{\zz}{\mathbb{Z}}

\newcommand{\disp}[1]{\displaystyle{#1}}

\newcommand{\eco}{\varphi_{\alpha}}

\newcommand{\rd}{\mathbb{R}^{d}}
\newcommand{\rdd}{\mathbb{R}^{2d}}
\newcommand{\supp}{\mbox{Supp}}
\newcommand{\tr}{\mbox{Tr}}

\newcommand{\detp}{\mbox{det}^{-\frac{1}{2}}_+ }

\def\qed{\hfill$\square$}

\newcommand{\ud}{\frac{1}{2}}

\newcommand{\wt}{\widehat{W_t}}
\newcommand{\wtu}{\widehat{W_1}}
\newcommand{\wtd}{\widehat{W_2}}

\newcommand{\ope}{op\'erateur }

\newcommand{\nrjl}{\Sigma_{\lambda}}
\newcommand{\grad}{\nabla H(z)}
\newcommand{\jgrad}{J\nabla H(z)}

\newcommand{\ceg}{\mathcal{C}_{E,g}}

\newcommand{\carb}{ \overline{\chi (g)} }
\newcommand{\mg}{M(g)}
\newcommand{\mgm}{M(g^{-1})}
\newcommand{\gmu}{g^{-1}}
\newcommand{\la}{\lambda}
\newcommand{\e}{\varepsilon}
\newcommand{\aaa}{\alpha}
\newcommand{\gmm}{\gamma}

\newcommand{\nrj}{\Sigma_E}

\newcommand{\vfi}{\varphi}

\newcommand{\lde}{L^2(\mathbb{R}^d)}
\newcommand{\ldec}{L^2_{\chi}(\mathbb{R}^d)}

\newcommand{\op}{Op_h^w}
\newcommand{\mm}{\tilde{M}(g)}
\newcommand{\mmu}{\tilde{M}(g)^{-1}}
\newcommand{\hq}{\widehat{H}}
\newcommand{\hqr}{\widehat{H}_{\chi}}
\newcommand{\chg}{\Lambda_h}
\newcommand{\trp}{\mathcal{T}_h(\alpha)}
\newcommand{\trpt}{\mathcal{T}_h(\alpha_t)}

\newcommand{\gr}{\int_{\mathbb{R}}}

\newcommand{\gra}{\int_{\mathbb{R} ^{2d}_{\alpha}} }

\begin{document}
\title[Reduced Gutzwiller formula with symmetry: case of a finite group]
{Reduced Gutzwiller formula with symmetry: \\ case of a finite group}
\author[Roch Cassanas]{Roch Cassanas}
\address{Laboratoire de Math\'ematiques Jean Leray\\
         Universit\'e de Nantes\\
         2, Rue de la Houssini\`ere, BP92208\\    
         F-44322, Nantes Cedex 03, France}
     \email{cassanas@math.univ-nantes.fr}
%
\begin{abstract}
\noindent 
We consider a classical Hamiltonian $H$ on $\rdd$, invariant by a finite group of symmetry $G$, whose Weyl
quantization $\hq$ is a selfadjoint operator on $\lde$. If $\chi$ is an irreducible character of $G$, we
investigate the spectrum of its restriction $\hqr$ to the symmetry subspace $\ldec$ of $\lde$ coming from
the decomposition of Peter-Weyl.
We give reduced semi-classical asymptotics of a regularised spectral density
describing the spectrum of $\hqr$ near a non critical energy $E\in\rr$. If $\nrj:=\{ H=E \}$ is compact,
assuming that periodic orbits are non-degenerate in $\nrj/G$, we get a reduced Gutzwiller trace formula
which makes periodic orbits of the reduced space $\nrj/G$ appear. The method is based upon the use of
coherent states, whose propagation was given in the work of M. Combescure and D. Robert.

\end{abstract}
\maketitle

\section{Introduction}
The purpose of this work is to give a Gutzwiller trace formula for a reduced quantum Hamiltonian in the
framework of symmetries given by a finite group $G$ of linear applications of the configuration space $\rd$. This semi-classical trace formula
 will link the reduced spectral density to periodic orbits of the dynamical system in the classical reduced
space, i.e. the space of $G$-orbits.\footnote{Results of this paper were published without proof in a Note
aux Comptes Rendus (see \cite{cras1}).}\

The role that symmetry plays in quantum dynamics was obvious since the beginning of the theory, and
emphasized by Hermann Weyl in the book: `The theory of groups and quantum mechanics' (\cite{W}).
Pioneering physical results
were given for models having a lot of symmetries. In the mathematical domain, first systematical
investigations were done in 1978-79, mainly for the eigenvalues counting function of the Laplacian on a
Riemannian compact manifold simultaneously by Donnelly and Br\"uning \& Heintze (see \cite{B-H} and
\cite{Don}). Later, Guillemin and Uribe described the relation with closed trajectories in \cite{G-U 1} and
\cite{G-U 2}.
In $\rd$, a general study was done in the early 80's for globally elliptic pseudo-differential operators, both in cases
of compact finite and Lie groups, by Helffer and Robert (see \cite{He-Ro 2}, \cite{He-Ro 3}) for high energy asymptotics, and later by
El Houakmi and Helffer in the semi-classical setting (see \cite{El.H}, \cite{El.H-He}). Main results were then given in terms of reduced
asymptotics of Weyl type for a counting function of eigenvalues of the operator.
Here, in a semi-classical study with a finite group of symmetry, we want to go one step beyond Weyl
formulae, investigating oscillations of the spectral density, and establishing a Gutzwiller
formula for the {\it reduced} quantum Hamiltonian. The case of a compact Lie group will be carried out in
another paper (see \cite{grouplie} and \cite{these}).\

Without symmetry, in 1971, M.C. Gutzwiller published for the first time his trace formula linking
semi-classically the spectrum of a quantum Hamiltonian $\hq$ near an energy $E$, to periodic orbits of the
classical Hamiltonian system of $H$ on $\rdd$, lying in the energy shell $\nrj:=\{ H=E \}$. This was one of
the strongest illustrations of the so-called `correspondence principle'. Later, rigorous mathematical proofs
were given (see for example \cite{P-U}, \cite{P-U2}, \cite{Mein}, \cite{Do}), using various techniques
like wave equation, heat equation, microlocal analysis, and more recently wave packets (see \cite{CRR}).\

Coming back to classical dynamics, let $H:\rdd\to\rr$ be a smooth Hamiltonian with a
finite group of symmetry $G$, such that $H$ is $G$-invariant, i.e. suppose that there is an action $M$ from
$G$ into $Sp(d,\rr)$, the group of symplectic matrices of $\rdd$, such that:
\beq\label{Ginvariant}
H(M(g)z)=H(z), \qquad \forall g\in G, \quad \forall z\in\rdd.
\eeq 
The Hamiltonian system associated to $H$ is:
\beq\label{hamsyst}
\dot{z}_t=J\nabla H(z_t), \mbox{ where } J=\left(
\barr{cc}
0    & I_d\\
-I_d & 0
\earr\right).
\eeq
In the framework of symmetry, specialists in classical dynamics are used to investigate this system in the space of $G$-orbits : $\rdd/G$, also called the
{\it reduced space}.\

Here, for a quantum study with symmetry, it is therefore natural to expect a reduced
Gutzwiller formula, linking semi-classically the spectrum of the {\it reduced} quantum Hamiltonian near the energy
$E$ to periodic orbits of the {\it reduced} classical dynamical system on $\nrj/G$.\\\\
We now briefly describe our main result. First, we introduce our quantum
reduction. We follow the same setting as in articles of Helffer and Robert \cite{He-Ro 2}, \cite{He-Ro 3}:
let $H:\rdd\to \rr$ be a smooth Hamiltonian and $G$ a finite subgroup of the
linear group $Gl(d,\rr)$. If $g\in G$, we set:
\beq\label{action}
M(g)(x,\xi):=(g\, x,^tg^{-1}\xi)
\eeq
and we assume that $H$ is $G$-invariant as in (\ref{Ginvariant}).
As usual, we make suitable assumptions -see (\ref{hypCF})- to have nice properties for the Weyl quantization
of $H$ (as functional calculus), which is defined as follows : for $u\in \mathcal{S}(\rd)$,
\beq\label{pseudoweyl}
\op(H) u(x)=(2\pi h)^{-d}\int_{\rd}\int_{\rd}e^{\frac{i}{h}(x-y)\xi}
H\left(\frac{x+y}{2},\xi\right) u(y) dy d\xi.
\eeq
In particular, $\op(H)$ is essentially selfadjoint on $\mathcal{S}(\rd)$ and we denote by $D(\hq)$, $\hq$
its selfadjoint extension.\

$G$ acts on the quantum space $\lde$ by $\tilde{M}$ defined for $g\in G$ by  :
\beq\label{actionquantique}
\mm(f) (x)=f(\gmu x), \qquad \forall f\in\lde,\quad \forall x\in\rd.
\eeq
If $\chi$ is an irreducible character of $G$, we set $d_{\chi}:=\chi(Id)$. Then, we define the symmetry
subspace $\ldec$ associated to $\chi$, by the image of $\lde$ by the projector:
\beq\label{projo}
P_{\chi}:=\frac{d_{\chi}}{|G|}\sum_{g\in G} \carb \mm,
\eeq
$\lde$ splits into a Hilbertian sum of $\ldec$'s (Peter-Weyl decomposition), and the property (\ref{Ginvariant}) implies that each
$\ldec$ is stable by $\hq$. Our goal is to give semi-classical trace formulae for the restriction $\hqr$ of
$\hq$ to $\ldec$, which will be called the {\it reduced quantum Hamiltonian}. We define the following 
 {\it reduced regularized spectral density} :
\beq\label{densitespectrale}
\mathcal{G}_{\chi}(h):=\tr\left(\psi(\hq_{\chi}) f\left( \frac{E-\hq_{\chi}}{h} \right) \right),
\eeq
where $\psi$ is smooth, compactly supported in a neighbourhood $]E-\delta E, E+\delta E[$ of $E\in\rr$
($\delta E>0$) such that
$H^{-1}([E-\delta E, E+\delta E])$ is compact ($\psi(\hqr)$ is an energy cut-off which is trace class),
$f$ is smooth and $\hat{f}$ (the Fourier transform of $f$) is compactly supported in $\rr$. The case where
$\supp(\hat{f})$ is  localised near zero is the one that leads to Weyl formulae, and gives an asymptotic
expansion  of the counting function of $\hqr$ (see Theorem \ref{Ig_W}, Corollary \ref{comptage}). Here we want to focus on the oscillating part of
$\mathcal{G}_{\chi}(h)$. Thus we suppose that $0\notin \supp(\hat{f})$.\

In order to state the theorem in terms of the reduced space, we need a smooth structure on $\nrj/G$, and
thus we suppose that the group acts freely on $\nrj$, so that dynamics of $H$ on $\nrj$
would descend to the quotient. Note that this is
not an essential assumption, since we have proved the asymptotic without this hypothesis (see Theorem
\ref{Gutz_osc}).
The following result involves the quantity $\chi(g_{\overline{\gmm}}^n)$, defined as follows: if $\pi$ denotes the projection on the quotient and $\overline{\gmm}$ is a periodic orbit in
$\nrj/G$, if $\pi(\gmm)=\overline{\gmm}$, then, there is only one $g_{\gmm}$ in $G$ such that,
$\forall z\in\gmm$, $M(g_{\gmm})\Phi_{T_{\overline{\gmm}}^*}(z)=z$, where $T_{\bar{\gmm}}^*$ is the primitive period
of $\bar{\gmm}$. If $\pi(\gmm_1)=\pi(\gmm_2)$ then
$g_{\gmm_1}$ and $g_{\gmm_2}$ are conjugate elements of $G$, and we denote by $\chi(g_{\overline{\gmm}})$ the
quantity $\chi(g_{\gmm_1})=\chi(g_{\gmm_2})$.\

In order to have a finite number of periodic orbits of the reduced space involved in the trace formula, we
will suppose that periodic orbits of $\nrj/G$ are {\it non-degenerate}, in the following sense :
If $\bar{\gmm}$ is a periodic orbit of $\nrj/G$, with primitive period $T_{\overline{\gmm}}^*$, and if
$n\in\zz^*$ is such that $nT_{\bar{\gmm}^*}\in\supp(\hat{f})$, then $1$ is not an eigenvalue of the
differential of the Poincar\'e map in $\nrj/G$ at $nT_{\bar{\gmm}^*}$: 
$\ker[(dP_{\overline{\gmm}})^n-Id]=\{ 0 \}$.
Then we have the following result:
\begin{theo}\label{Gutzreduit}
Under previous assumptions, suppose that the group $G$ acts freely on $\nrj$ and that periodic orbits of
$\nrj/G$ are non-degenerate in the sense given above. We then
have a complete asymptotic expansion of $\mathcal{G}_{\chi}(h)$ in powers of $h$, modulo an oscillating
factor of the form $e^{i\frac{\aaa}{h}}$  as $h\to 0^+$ (see Theorem \ref{Gutz_osc} for  details). The first term is given by:
$$\mathcal{G}_{\chi}(h)= d_{\chi}\psi(E)
\sum_{\tiny{\barr{c}
\overline{\gmm}  \mbox{ periodic }\\
\mbox{orbit of } \nrj/G
\earr}}
\sum_{\tiny{\barr{c}
n\in\zz^*  \mbox{ s.t. }\\
nT_{\overline{\gmm}}^*\in \supp \hat{f}
\earr}}
\hat{f}(nT_{\overline{\gmm}}^*) \overline{\chi(g_{\overline{\gmm}}^n)}
e^{\frac{i}{h} n S_{ \overline{\gmm} }}\frac{ T_{\overline{\gmm}}^* \; e^{i\frac{\pi}{2}\sigma_{\overline{\gmm},n}}}
{2\pi |\det((dP_{\overline{\gmm}})^n-Id)|^{\ud}}
+O(h).$$
where $S_{ \overline{\gmm}}:=\int_0^{T_{\overline{\gmm}}^*} p_s \dot{q}_s ds$, $P_{\overline{\gmm}}$ is the Poincar\'e
map of $\overline{\gmm}$ in $\nrj/G$, and $\sigma_{\overline{\gmm},n}\in \zz$. The other terms are distributions
in $\hat{f}$, with support in the set of periods of orbits in $\nrj/G$.
\end{theo}
\textsl{Remark 1}: the case with $0\in\supp(\hat{f})$ could have been included in the preceding theorem, and we
would get a Weyl term in addition to this oscillating part. This term was already described by El Houakmi
(see \cite{El.H}) for the leading contribution. We obtain here slightly more detailled asymptotics for the
Weyl part, by calculating the contribution of each $g\in G$ : see Theorem \ref{Ig_W}.\\\

\textsl{Remark 2}: one could also consider a symmetry directly given in phase space $\rd\times\rd$, and set
$G$ as a finite subgroup of $Sp(d,\rr)$. Then we would have to suppose that there is a unitary action
$\tilde{M}:G\to\mathcal{L}(\lde)$ which is metaplectic, i.e. satisfies:
\beq\label{metaplectoc}
\mm^{-1}\op(H)\mm=\op(H\circ g),\, \mbox{ for all $g$ in $G$}.
\eeq
For a fixed $g$, there is always some $\mm$ satisfying (\ref{metaplectoc}), but it is not unique (multiply $\mm$ by a
complex of modulus $1$). The difficulty is to find a $\tilde{M}$ that is also a group homomorphism.\\\

The method used is close to the one of \cite{CRR} : unlike articles previously quoted, which used an
approximation of the propagator $\exp(-i\frac{t}{h}\hq)$ by some FIO following the WKB method, we will use
here the work of Combescure and Robert on the propagation of coherent states. This method avoids problems of
caustics and looks simpler to us. Moreover, the symmetry behaves well with coherent states, and we get
very pleasant formulae. Thanks to these wave packets, we first reduce the problem to an
application of the generalised stationary phase theorem (section \ref{reduc_ec}). Then we find minimal hypotheses for the critical set to be a smooth manifold, and to
ensure that the transverse Hessian of the phase is non-degenerate. These hypotheses will be called
`$G$-clean flow conditions', and we get a theorical asymptotic expansion of $\mathcal{G}_{\chi}(h)$ under these
assumptions (Theorem \ref{flotpropre}).
Finally, as particular cases, we will show that these conditions are fulfilled on the one hand when $\hat{f}$
is supported near zero (`Weyl term' Theorem \ref{Ig_W}), and on the other hand when periodic
orbits are non-degenerate (`Oscillating term' Theorem \ref{Gutz_osc}). In both cases, we calculate
geometrically first terms of the asymptotic expansion, to make quantities of the reduced classical dynamics
appear, as the energy level, periodic orbits and the Poincar\'e map. The symmetry of periodic orbits plays
an important part in the result.\\\
 
\textbf{Aknowledgements}: We found strong motivation in the work of physicists B. Lauritzen, J.M. Robbins,
and N.D. Whelan (\cite{L}, \cite{L-W}, \cite{R}). I am deeply grateful to Didier Robert for his help,
comments and suggestions. Part of this work was made with the support received from the ESF (program SPECT).
I also thank Ari Laptev for many stimulating conversations.
\section{Details on quantum reduction}

\subsection{Symmetry subspaces} We recall some basic facts on representations (see \cite{Se}, \cite{Si} or
\cite{Pi}): a representation $\rho : G\to Gl(E)$ of the group $G$ on a
finite dimensional complex vector space $E$ is said to be irreducible if there is no non-trivial subspace
of $E$ stable by $\rho(g)$, for all $g$ in $G$. The character $\chi_\rho:G\to \cc$ of a
representation is defined by $\chi_{\rho} (g):=Tr(\rho(g))$, for $g\in G$. The degree of the representation
$\rho$ is denoted by $d_{\chi_{\rho}}$ and is the dimension of $E$. Two such representations are isomorphic
if and only if they have the same character. We will denote by $\widehat{G}$ the set of all irreducible
characters, that is the set of characters of irreducible representations. Moreover, $G$ finite implies $\widehat{G}$
finite.\

A representation $\tilde{M}$ of $G$ on a Hilbert space is said to be unitary if each $\tilde{M}(g)$ is a unitary
operator. This is the case of our representation $\tilde{M}$ on the Hilbert space $\lde$ defined by
(\ref{actionquantique}) since $|\det(g)|=1$. One can easily check that $\tilde{M}$ is strongly continuous. Then, the Peter-Weyl
theorem (see \cite{Si} or \cite{Pi}) says that if one set $\ldec:=P_\chi(\lde)$, where $P_\chi$ is defined by
(\ref{projo}), then the $P_\chi$'s are orthogonal projectors of sum identity, and we have the Hilbertian
decomposition:
\beq
L^2(\rr^d)=   \bigoplus_{ \mbox{ {\scriptsize $\chi\in\widehat{G}$ } }}^{\perp}  \quad \ldec .
\eeq
Furthermore, if $\chi\in\widehat{G}$, then any irreducible sub-representation of $\tilde{M}$ in $\ldec$ is
of character $\chi$, and a decomposition having such a property is unique. These $\ldec$'s will be called here the {\it symmetry subspaces}.\

One has to think of them as a certain class of functions of $\lde$ having a certain symmetry linked to $G$
and $\chi$. For example, if $G=\{ \pm Id_{\rd} \}$, then we have two irreducible characters $\chi_+$ and
$\chi_-$ such that $L^2_{\chi_+}(\rd)$ is the set of even functions of $\lde$, and $L^2_{\chi_-}(\rd)$ is the set of
odd functions. More generally, if $\chi$ is a character of degree $1$, then $\chi$ is
multiplicative, and we have:
$$\ldec=\{ f\in\lde\; : \;\forall g\in G,\; \mm f=\chi(g) f\, \}.$$
This is in particular the case for abelian groups. If $G\simeq$ {\Large $\sigma_d$} is the symmetric group of permutation matrices
acting on $\rd$, then there is at least two characters of degree $1$: $\chi_0$, the trivial character
(always equal to $1$), and the signature $\varepsilon$. Thus we get:
\begin{itemize}
\item[-] $L^2_{\chi_0}(\rd)=\{ f\in\lde\; : \; \forall \sigma\in G,\;
f(x_{\sigma(1)},\dots,x_{\sigma(d)})=f(x_1,\dots,x_d) \}$.
\item[-] $L^2_{\varepsilon}(\rd)=\{ f\in\lde\; : \; \forall \sigma\in G,\;
f(x_{\sigma(1)},\dots,x_{\sigma(d)})=\varepsilon (\sigma) f(x_1,\dots,x_d) \}$.
\end{itemize}

\subsection{Reduced Hamiltonians} It is easy to check on the formula (\ref{pseudoweyl}) that we have on
$\mathcal{S}(\rd)$:
\beq\label{metaplec}
\mmu \op(H)\, \mm=\op(H\circ \mg), \quad \forall g\in G.
\eeq
Thus we see that the property of $G$-invariance (\ref{Ginvariant}) is equivalent to the commutation of $\hq$
with all $\mm$. In particular, it implies that $\hq$ commutes with all $P_\chi$'s, and thus, $\ldec$ is stable
by $\hq$. We can then define the operator that we plan to study: if $\chi\in\widehat{G}$, set:
$$D(\hqr):=\ldec\cap D(\hq),$$
The restriction  $\hqr$  of $\hq$ to $\ldec$ is called the {\it reduced quantum Hamiltonian}, and is a
selfadjoint operator on the Hilbert space $\ldec$. If $f:\rr\to \cc$ is borelian, then we have:
$$[f(\hq),P_\chi] =0, \quad D(f(\hqr))=D(f(\hq))\cap \ldec,\quad f(\hq)=\sum_{\chi\in\widehat{G}}f(\hqr) \, P_\chi$$
$f(\hqr)$ is the restriction of $f(\hq)$ to $\ldec$.
Lastly, if $\sigma(.)$ denotes the spectrum of
an operator, then we have: $\disp{ \sigma(\hq)=\underset{\chi\in\widehat{G}}{\cup} \sigma(\hqr) }$ (for details, see
\cite{these}).\\\

One trace formula will be essential for the rest of this article:
\begin{lem}
If $f:\rr\to\rr$ is borelian, and if $f(\hq)$ is trace class on $\lde$, then, for all $\chi\in\widehat{G}$, $f(\hqr)$
is trace class on $\ldec$ and:
\beq\label{tracered}
\tr(f(\hqr))=\tr(f(\hq) P_\chi).
\eeq
\end{lem}
Indeed, we have to show that $|f(\hqr)|^{\ud}$ is Hilbert-Schmidt and
$\norm{|f(\hqr)|^{\ud}}_{HS}\leq \norm{|f(\hq)|^{\ud}}_{HS}$,
which is clear by completing an Hilbertian basis of $\ldec$ in an Hilbertian basis of $\lde$. Then one writes:
$$Tr(f(\hqr))=\sum_{\la\in\sigma(f(\hqr))-\{ 0 \}}\dim (Ker[f(\hqr)-\la])\, \la$$
$$Tr(f(\hq)\circ P_\chi)=\sum_{\la\in\sigma(f(\hq)\circ P_\chi)-\{ 0 \}}\dim (Ker[f(\hq)\circ P_\chi-\la])\la .$$
Furthermore, if $\la\neq 0$, then $Ker(f(\hq)\circ P_\chi-\la)=Ker(f(\hqr)-\la)$, and we get (\ref{tracered}).
\subsection{Interpretation of the symmetry}
The investigation of $\hqr$ provides informations on the spectrum of $\hq$:
\begin{lem}
If $\chi\in \widehat{G}$ then eigenvalues of $\hqr$ have a multiplicity proportional to $d_\chi$.
\end{lem}
Indeed, if  $F\subset \ldec$ is an eigenspace of $\hqr$, then it is $\tilde{M}$-invariant. One
can decompose it into irreducible representations. By the Peter-Weyl theorem, the only irreducible
representation appearing is the one of character $\chi$, and thus is of dimension $d_\chi$.
In particular, the operator $\hqr$ provides a lower band for the multiplicity of some eigenvalues of $\hq$.\

Another remark: by splitting an eigenfunction of $\hq$ on the symmetry subspaces, we get at least an
eigenvector in one $L^2_{\chi}(\rd)$. This means that each eigenspace of $\hq$ contains an eigenvector having a
certain symmetry. As it is well know for the double well potential ($G=\{ \pm Id \}$), where eigenspaces are of dimension $1$,
this leads to an alternance of even/odd eigenspaces and to tunneling effect.\

If $N_\chi(I)$ denotes the number of eigenvalues of $\hqr$ (with multiplicity) in an interval $I$ of $\rr$,
and $N(I)$ the one of $\hq$, then the quantity $N_\chi(I)/N(I)$ can be thought as the proportion of
eigenfunctions of symmetry $\chi$ among those corresponding to eigenvalues of $\hq$.
\subsection{Examples}
We give a few examples of Schr\"odinger Hamiltonians with a finite group of symmetry:
$$H(x,\xi):=|\xi|^2+V(x).$$
\begin{enumerate}
\item $G=\{ \pm Id \}$ : double well: $V(x)=(x^2-1)^2$, harmonic or quartic oscillator: $V(x)=x^2$ or $x^4$,
`the well on the island': $V(x)=(x^2+a)e^{-x^2}$ ($a>0$). For the two first examples,
$V(x)\xrightarrow[+\infty]{}+\infty$, so $\hq$ is essentially selfadjoint on $\mathcal{S}(\rr)$ and
with compact resolvant.\\
\item $G\simeq${\Large $\sigma_2$}, $d=2$: any potential satisfying $V(x,y)=V(y,x)$.\\
\item Group of isometries of the triangle, $d=2$: $V(x,y)=\ud(x^2+y^2)^2-xy^2+\frac{1}{3}x^3$, which in polar
coordinates is $\tilde{V}(r,\theta)=V(r\cos \theta,r\sin \theta)=\ud r^2+\frac{1}{3}r^3\cos(3\theta)$ (see
also the H\'enon-Heiles potential: $V(x,y)=\ud(x^2+y^2)-xy^2+\frac{1}{3}x^3$, but one has to look for the
selfadjointness of this operator).\\
\item Group of isometries of the square, $d=2$: $V(x,y)=\ud x^2y^2$.\\
\item $G\simeq (\zz/2\zz)^d$: harmonic oscillator with distinct frequencies: $V(x)=<Sx,x>_{\rd}$, with $S$
symmetric positive definite matrix with eigenvalues pairwise distincts. In this case, $\hq$ is still
essentially selfadjoint  on $\mathcal{S}(\rd)$ and with compact resolvant. This is one of the few cases where
we can calculate periodic orbits of the dynamical system.
\end{enumerate}
\section{Reduction of the proof by coherent states }\label{reduc_ec}
We adapt here the method of \cite{CRR}. The essential tool is the use of coherent states.\footnote{More
details on the proof can be found in \cite{these}.} We
refer to the Appendix where we recall basic things about it (se also \cite{CR}, \cite{CRR}, or \cite{these}).
Note that, by an averaging argument (see section \ref{Weylpart}), we could already restrict ourselves to a
group of {\it isometries}. For the moment, we still use the general expression of (\ref{action}), to keep in mind
the symplectic form of $M(g)$. We suppose that $\psi$ and $f$ are in $\mathcal{S}(\rr)$ such that
$\supp(\psi)\subset ]E-\delta E, E+\delta E[$ and the Fourier transform $\hat{f}$ of $f$ is  with compact
support. We know from \cite{He-Ro 1}, \cite{Ro}, that, under hypothesis (\ref{hypCF}), $\psi(\hq)$ is trace class for little $h$'s, and, by formula (\ref{tracered}), we have:
$$\mathcal{G}_\chi (h)=Tr\left(\psi(\hqr) f\left( \frac{E-\hqr}{h} \right)\right)=\frac{d_{\chi}}{|G|}\sum_{g\in G}\carb
I_g(h),$$
where:
\beq\label{Ig}
I_g(h):=Tr\left(\psi(\hq) f\left( \frac{E-\hq}{h} \right)\mm\right).
\eeq
Then, by Fourier inversion, we make the $h$-unitary quantum propagator $U_h(t):=e^{-i\frac{t}{h}\hq}$ appear,
and write:
\begin{equation}\label{Ig1}
I_g(h)= \frac{1}{2\pi}\gr e^{i\frac{tE}{h}}.
\hat{f}(t).\mbox{Tr}\left(\psi(\hq)U_h(t)\mm\right)dt.
\end{equation}
Then we use the trace formula with coherent states -- see (\ref{trace_ec}) -- to write:
\begin{equation}\label{chloe}
I_g(h)=\frac{(2\pi h)^{-d}}{2\pi} \gr \gra e^{i\frac{tE}{h}}.\hat{f}(t).m_h(\aaa,t,g)d\aaa dt.
\end{equation}
where
\begin{equation}\label{marine}
m_h(\aaa,t,g):=<U_h(t)\eco;\mmu\psi(\hq)\eco>_{\lde}.
\end{equation}
With exactly the same proof as in \cite{CRR}, we get the following lemma:
\begin{lem}\label{compactenalpha}
There exists a compact set $K$ in $\rdd$ such that:
$$\int_{\rdd\setminus K}|m_h(\aaa,t,g)| d\aaa= O (h^{+\infty}).$$
uniformly with respect to $g\in G$ and $t\in\rr$.
\end{lem}
We can then suppose that $\nrj:=\{ H=E\}$ is included in $K$, and choose a real cut-off function $\chi_1$, compactly
supported in $\rdd$ and equal to $1$ on $K$. We can write $1=\chi_1+(1-\chi_1)$ in (\ref{chloe}), and  settle
problems at infinity in $\alpha$.
Besides, we want to use the functional calculus of Helffer and Robert (\cite{He-Ro 1}, \cite{Ro})
for the description of $\psi(\hq)$. Thus we make the following hypothesis:
$\exists C>0$, $\exists C_{\alpha}>0$, $\exists m>0$ such that:
\beq\label{hypCF}
\left\{
\begin{array}{l}
<H(z)>\leq C<H(z^{\prime})>.<z-z^{\prime}>^m, \quad \forall z,z^{\prime} \in \rr^{2d}.\\
|\partial_z^{\alpha} H(z)|\leq C_{\alpha}<H(z)>, \quad\forall z \in \rr^{2d}, \forall\alpha \in
\nn^{2d}.\\
H \mbox{ has a lower band on }\rdd.
\end{array}
\right.
\eeq
Then, we can write for $N_0\in\nn$:
\beq\label{hero2}
\psi(\hq)= \sum_{j=0}^{N_0} h^j \op(a_j) + h^{N_0+1}.R_{N_0+1}(h).
\eeq
where $\supp (a_j) \subset H^{-1}(]E-\delta E,E+\delta E[)$, $a_0(z)=\psi (H(z))$, with
$\disp{ \underset{0<h\leq 1}{ \mbox{Sup} } \norm{R_{N_0+1}(h)}_{\tr} \leq C.h^{-d} }.$\

We obtain:
\beq
I_g(h)=\sum_{j=0}^{N_0} h^j I_g^{j}(h) + O(h^{-d} h^{N_0+1}).
\eeq
Now, we must get a complete asymptotic expansion for a fixed $j_0$ in $\nn$ of the quantity:
\beq\label{Igzero}
I_g^{j_0}(h)=\frac{(2\pi h)^{-d}}{2\pi} \gr \gra e^{i\frac{tE}{h}}.\hat{f}(t) \chi_1(\aaa)\,
m_h^{j_0}(\aaa,t,g)d\aaa dt,
\eeq
with
\beq\label{cocodingo}
m_h^{j_0}(\aaa,t,g):=<U_h(t)\eco;\mmu\op(a_{j_0})\eco>_{\lde}.
\eeq
\newline
{\it For the right term of the bracket in (\ref{cocodingo})}, 
we expand $\op(a_{j_0})\eco$ in powers of $h$, by Lemma 3.1 of \cite{CRR}. Thanks to (\ref{metaplec}), since
$\trp=\op(\exp(\frac{i}{h}(px-q\xi))$ -- see Appendix -- we can write:
$$\mmu \mathcal{T}_h(\aaa)=\mathcal{T}_h(\mgm \aaa) \mmu.$$
{\it For the left term of the bracket in (\ref{cocodingo})}, we use the theorem of propagation of coherent
states given by Combescure and Robert (\cite{CR}, \cite{CRR} or \cite{Ro1}). If $M\in\nn$, then we have:
$$\norm{U_h(t)\eco-e^{i\frac{\delta(t,\aaa)}{h}} 
\trpt\chg \left[\sum_{j=0}^M h^{\frac{j}{2}} \, b_j(t,\aaa)(x).e^{\frac{i}{2}<M_0x,x>}\right] }_{\lde}
\leq C_{M,T}(\aaa).\, h^{\frac{M+1}{2}}$$
where $\aaa_t=\Phi_t(\aaa)$ is the solution of the system (\ref{hamsyst}) with initial condition $\aaa$
(see Appendix for other notations).
After all, since there is no problem of control for $\aaa$ at infinity, we get:
\beq\label{mh}
m_h^{j_0}(\aaa,t,g)=\sum_{k=0}^{2N}\sum_{j=0}^{2N-k} h^{\frac{j}{2}}h^{\frac{k}{2}}
\sum_{|\gmm|=k}\frac{ \partial^{\gmm}a_{j_0}(\aaa) }{\gmm !}
\, e^{i\frac{\delta(t,\aaa)}{h}}\,\Upsilon_{j,\gmm}(\aaa,t,g,h)
+O(h^{-d}h^{N+\ud}),
\eeq
with:
$$\Upsilon_{j,\gmm}(\aaa,t,g,h):=<\trpt\chg b_j(t,\aaa)e^{\frac{i}{2}<M_0x,x>};\mathcal{T}_h(\mgm \aaa)\chg
\mmu Q_{\gmm}\tilde{\psi}_0>$$
where $Q_\gmm$ is the polynomial in $d$ variables such that:
\beq
Op_1^w(z^{\gmm})\tilde{\psi}_0=Q_{\gmm}.\tilde{\psi}_0
\eeq
We have: $\chg^*
\mathcal{T}_h(-\mgm\aaa)\mathcal{T}_h(\aaa_t)\chg=e^{\frac{i}{2h}<\mgm\aaa,J\aaa_t>}\mathcal{T}_1\left(\frac{\aaa_t-\mgm\aaa}{\sqrt{h}}\right)$
(see Appendix). Thus:
$$\Upsilon_{j,\gmm}(\aaa,t,g,h)=e^{\frac{i}{2h}<\mgm\aaa,J\aaa_t>}
<\mathcal{T}_1\left(\frac{\aaa_t-\mgm\aaa}{\sqrt{h}}\right) b_j(t,\aaa)e^{\frac{i}{2}<M_0x,x>},\mmu Q_\gmm
\tilde{\psi}_0>_{L^2}.$$
We will use the notation:
$$\aaa=(q,p)\in\rd\times\rd \quad and \quad (q_t,p_t):=\aaa_t=\Phi_t(\aaa).$$
Make the change of variable: $\gmu y:=x-(q_t-\gmu q)/\sqrt{h}$ in the previous $<;>_{L^2}$. Since $G$ is compact, $|\det(g)|=1$, and we
obtain after calculation:
$$\Upsilon_{j,\gmm}(\aaa,t,g,h)=\pi^{-\frac{d}{4}}e^{\frac{i}{h}[\frac{qp+q_tp_t}{2}-^tgpq_t]+\frac{i}{2}|gq_t-q|^2}
\int_{\rd}e^{-\ud <Ay,y>+\beta y}Q_\gmm\left(y+\frac{gq_t-q}{\sqrt{h}}\right) b_j(\aaa,t)(y)dy,$$
where
\beq\label{Abeta}
A:=I-^tg^{-1}M_0\gmu, \quad and \quad \beta:=\frac{i}{\sqrt{h}}[(q-gq_t)+i(^tg^{-1}p_t-p)].
\eeq
Then we set:
$$Q_{\gmm}(x)=:\sum_{|\mu|\leq |\gmm|}\kappa_{\mu,\gmm} x^{\mu} \quad\mbox{ and }\quad
b_j(t,\aaa)(x)=:\sum_{|\nu|\leq 3j}c_{\nu,j}(t,\aaa) x^{\nu}$$
(where $c_{\nu,j}$ is smooth in  $t,\aaa$). For the same reasons as in \cite{CRR} (parity of $Q_\gmm$ and
$b_j(t,\aaa)$), only entire powers of $h$ have non-zero coefficients. Then, we can expend $Q_\gmm$ and
$b_j(t,\aaa)$ and use the following calculus of the Gaussian:
\begin{lem}\label{gauss}
Let $A\in M_d(\cc)$ such that $^tA=A$, and that $\Re A$ is a positive definite matrix, $\beta \in \cc^d$ and
$\aaa\in \nn^d$. Then $A$ is invertible and
$$\int_{\rr^d_x} e^{-\frac{1}{2}<Ax,x>+\beta x}x^{\aaa}dx=(2\pi)^{\frac{d}{2}}\detp (A) e^{\ud
<A^{-1}\beta,\beta>}\sum_{\eta\leq \aaa}(A^{-1}\beta)^{\eta}P_{\eta}(A),$$
where $P_{\eta}(A)$ doesn't depend on $\beta$, and $P_0(A)=1$ (for a precise definition of $\detp$, see \cite{CRR}).
\end{lem}
We get: $\; e^{i\frac{t}{h}E}e^{\frac{i}{h}\delta(t,\aaa)}\; \Upsilon_{j,\gmm}(\aaa,t,g,h)=$
$$\sum_{|\nu|\leq 3j}\sum_{|\mu|\leq |\gmm|}
\kappa_{\mu,\gmm} c_{\nu,j}(t,\aaa) \sum_{\eta\leq \mu}\left(
\barr{c}
\mu\\
\nu
\earr
\right) (2\pi)^{\frac{d}{2}}\detp(I-i^t\gmu M_0\gmu) \sum_{\sigma\leq\mu-\eta+\nu}(gq_t-q)^{\eta}$$
\hfill $\disp{\times \left[(I-i^t\gmu M_0\gmu)^{-1}(\beta_0)\right]^{\sigma}
 P_\sigma(A) \, h^{-\ud(|\sigma|+|\eta|)} \exp\left(\frac{i}{h}\vfi_E(t,\aaa,g)\right) },$\\
 where $\beta_0:=\sqrt{h}\beta$, and:
 \beq\label{phaseboy}
 \vfi_E(t,\aaa,g)=tE+S(t,\aaa)+qp-^tgpq_t+\frac{i}{2}|gq_t-q|^2-\frac{i}{2}<A^{-1}\beta_0,\beta_0>.
 \eeq
Thus, (\ref{Igzero}) and (\ref{cocodingo}) give:
$$I_g^{j_0}(h)=\frac{(2\pi h)^{-d}}{2\pi}\sum_{k=0}^{2N}\sum_{j=0}^{2N-k} h^{\frac{j+k}{2}}
\sum_{|\gmm|=k} \sum_{|\nu|\leq 3j}\sum_{|\mu|\leq |\gmm|}\frac{\kappa_{\mu,\gmm} }{ \pi^{\frac{d}{4}}\gmm !} 
(2\pi)^{\frac{d}{2}}\sum_{\eta\leq \mu}\left(
\barr{c}
\mu\\
\nu
\earr
\right)L_{\eta,\nu,\mu,\gmm,j}(h) \; +\; O(h^{N+\ud-d}),$$
with:
\beq\label{Lnu}
L_{\eta,\nu,\mu,\gmm,j}(h):=\sum_{\sigma\leq\mu-\eta+\nu} h^{-\ud(|\sigma|+|\eta|)}
\int_{\rr_t}\int_{\rdd_\aaa} \exp\left(\frac{i}{h}\vfi_E(t,\aaa,g)\right) \hat{f}(t)\partial^{\gmm}a_{j_0}(\aaa)
\chi_1(\aaa)D_{\sigma,\eta,\nu,j}(t,\aaa,g)d\aaa dt.
\eeq
where: $D_{\sigma,\eta,\nu,j}(t,\aaa,g):=$
\beq\label{D}
c_{\nu,j}(t,\aaa) \detp(I-i^t\gmu M_0\,\gmu)P_\sigma(A) 
\left[ A^{-1}[(q-gq_t)+i(^tg^{-1}p_t-p)]\right]^{\sigma} (gq_t-q)^{\eta}.
\eeq
A tiresome but straightforward computation gives from (\ref{phaseboy}) and (\ref{Abeta}):
$$\vfi_E=\vfi_1+i \vfi_2.$$
\beq\label{phase1}
\left\{
\barr{l}
\disp{ \vfi_1(t,\aaa,g):=(E-H(\aaa))t+\ud<M(g)^{-1}\aaa,J\aaa>-\ud\int_0^t(\aaa_t-\mgm\aaa)J\dot{\aaa_s}ds}\\
\disp{ \vfi_2(t,\aaa,g):=\frac{i}{4}<(I-\wt)(\mg\aaa_t-\aaa);(\mg\aaa_t-\aaa)>.}
\earr\right.
\eeq
where
$\widehat{W_t}:=\left(
\barr{cc}
W_t      & -iW_t\\
-i\, W_t & -W_t
\earr
\right)$
with $\ud (I+W_t):=(I-i^tg^{-1}\, M_0\gmu)^{-1}$.\\
\begin{lem}\label{normWt}
We have: $\norm{W_t}_{\li(\cc^{d})}<1.$
\end{lem}
\underline{{\it Proof}}: we introduce the Siegel half-plane:
$$\Sigma_d:=\{Z\in M_d(\cc) : ^tZ=Z, \mbox{ and } \Im Z \mbox{ is positive definite }\}.$$
We know from \cite{Fo} pp.202, 203 that if $Z\in\Sigma_d$, then  $\norm{(I-iZ)^{-1}(I+iZ)}_{\li(\cc^d)}<1$.
Now, we can take $Z=^tg^{-1}\, M_0\gmu$. Indeed $M_0$ is symmetric, and, since $F_\aaa(t)$ is symplectic,
we have:
$$\forall X\in\rr^d,\qquad \Im (^tX.M_0 X)=|(A+iB)^{-1}X|_{\cc^d}^2.$$
Thus $Z\in\Sigma_d$. The proof is clear if we note that
$(I-i^tg^{-1}\,M_0\,\gmu)^{-1}(I+i^tg^{-1}\,M_0\,\gmu)=W_t$.$\square$\\\

We are led to solve a stationary phase problem to get an expansion of each
$L_{\eta,\nu,\mu,\gmm,j}(h)$ in powers of $h$.\\\\
\textsl{Remark}: Note that the term $D_{\sigma,\eta,\nu,j}$ -- (\ref{D}) -- and its derivatives will be vanishing on the
critical set of the phase for derivatives up to $|\sigma|+|\eta|$ (see (\ref{D}) and (\ref{critik})). Therefore, the
asymptotic of $\int_{\rr_t}\int_{\rdd_{\aaa}}\dots dtd\aaa$ will be shifted of $h$ to the power
$\ud(|\sigma|+|\eta|)$. This fact compensates for the term in $h^{-\ud(|\sigma|+|\eta|)}$, at the beginning of the expression of
$L_{\eta,\nu,\mu,\gmm,j}(h)$ in (\ref{Lnu}).
\section{The stationary phase problem}
Now, we fix $g$ in $G$ and we want to find the conditions under which we will be able to apply the
stationary phase theorem under the form of \cite{CRR} (Theorem 3.3) on $L_{\eta,\nu,\mu,\gmm,j}(h)$.
A necessary and
sufficient condition will be called `$g$-clean flow'. Then we will give particular cases for which
this criterium is satisfied (see sections \ref{Weylpart} and \ref{Gutzpart}). Our method will first
consist in calculating the critical set of the phase $\vfi_E$ and its Hessian. Then we will calculate
the kernel of this Hessian, and, under assumption of smoothness of the critical set, we will describe
the conditions for this kernel to be equal to the tangent space of the critical set.
In this section, since $g$ is fixed in $G$, we will denote $\vfi_E(t,z,g)$ by $\varphi_{E,g}(t,z)$, for
$z\in\rdd$ and $t\in\rr$.
\subsection{Computations and $g$-clean flow}$\quad$\\\

$\bullet$ \underline{Computation of the critical set}\\
$$\mbox{Let }\ceg:=\{ a\in \rr\times \rr^{2d} : \Im (\varphi_{E,g}(a))=0,
\nabla\varphi_{E,g}(a)=0 \}.$$
\begin{prop}\label{calculgenscrit}
The critical set is:
\begin{equation}\label{critik}
\ceg=\{(t,z)\in \rr\times \rdd : z\in\Sigma_E,\;\mg \Phi_t(z)=z \}.
\end{equation}
where $(t,z)\mapsto \Phi_t(z)$ is the flow of the system (\ref{hamsyst}).
\end{prop}
\underline{{\it Proof}} :
$$\Im \vfi_E(t,z,g)=\Re \vfi_2(t,z,g)=\frac{1}{4}|z_t-\mgm z|^2-\frac{1}{4}\Re <\wt(\mg z_t- z);\mg z_t-z>_{\rdd}.$$
We note that, if $a$ and $b$ are in $\rd$, then:
$$<\widehat{W_t}(a,b);(a,b)>_{\rdd}=<W_t(a-ib);(a-ib)>_{\rd}.$$
Thus,
$$\Im \vfi_E(t,z,g)=0 \iff |z_t-\mgm z|^2=\Re <W_t\beta,\beta>_{\rd}=\Re <W_t \beta,\overline{\beta}>_{\cc^d},$$
where
$$\beta:=(gq_t-q)-i(^tg^{-1}p_t-p).$$
Therefore, by lemma \ref{normWt}, we have:
$\Im \vfi_E(t,z,g)=0 \iff \Phi_t(z)=\mgm z$.\\\\
--{\it Computation of the gradient of $\varphi_1$} :\\
$\left\{
\barr{c}
\partial_t\vfi_1(t,z,g)=E-H(z)-\ud<(z_t-\mgm z);J\dot{z}_t>\\
\nabla_z\vfi_1(t,z,g)=\ud(^t\mgm+^tF_z(t))J(z_t-\mgm z)\\
\earr
\right.$\\\\
--{\it Computation of the gradient of $\varphi_2$} :\\
$\left\{
\barr{c}
4\partial_t\vfi_2(t,z,g)=2<(I-\widehat{W_t})(\mg z_t- z);\mg \dot{z}_t>-
<\partial_t(\widehat{W_t})(\mg z_t- z);(\mg z_t-z)>\\
4\nabla_z\vfi_2(t,z,g)=2(^tF_z(t)^t \mg-I)(I-\widehat{W_t})(\mg z_t- z)-
^t[\partial_z(\widehat{W_t})(\mg z_t-z)](\mg z_t-z)\\
\earr
\right.$\\\\
Thus, we see that $(t,z,g)\in \ceg$ if and only if $\Phi_t(z)=\mgm z$ et $H(z)=E$.\qed\\\

$\bullet$ \underline{Computation of the Hessian $\mbox{Hess} \;\vfi_{E,g}(t,z)$}\\\\
We first need some formulae coming from the symmetry that will be helpful for the computation:
We recall that $F_z(t)=\partial_z(\Phi_t(z))$. By differentiating formula (\ref{Ginvariant}), we get:
\beq\label{grad}
\nabla  H(\mg z)=^t\mgm \nabla H(z),\qquad \forall z\in\rdd,\;\forall g\in G.
\eeq
This formula implies that we have also:
\beq\label{flowsym}
\Phi_t(\mg z)=\mg \Phi_t(z), \qquad \forall z\in\rdd,\;\forall g\in G,\; \forall t\in\rr \mbox{ such that the flow exists at time $t$}.
\eeq
Moreover we recall that, since $\mg$ is symplectic, we have:
\beq
J\mg=^t\mgm J \;\mbox{ and }\; \mg J =J^t\mgm.
\eeq
Finally, if $t$ and $z$ are such that $\mg \Phi_t(z)=z$, then we have:
\beq\label{vecpropre}
(\mg F_z(t)-I)\jgrad=0 \mbox{ and } (^tF_z(t)^t\mg -I)\grad=0.
\eeq
The second identity comes from the first since $\mg F_z(t)$ is symplectic. For this first relation, one can
differentiate at $s=t$ the equation:
$$\Phi_t(\mg\Phi_s(z))=\Phi_s(z).$$
With these formulae, it is easy to find that:
\begin{prop}\label{calculhess}
$\mbox{Hess} \;\vfi_{E,g}(t,z)=$
$$ \left(
\barr{c|c}
\frac{i}{2}<(I-\wt) \jgrad;\jgrad> & -^t\nabla H(z)\\
 & + \frac{i}{2}^t\left[(^tF_z(t)^t\mg-I)(I-\wt) \jgrad \right] \\\hline

-\nabla H(z) & \ud[J\mg F_z(t)-^t(\mg F_z(t))J] \\
+\frac{i}{2}(^tF_z(t)^t\mg-I)(I-\wt) \jgrad & +\frac{i}{2}(^tF_z(t)^t\mg-I)(I-\wt)(\mg F_z(t)-I)\\
\earr
\right).$$
\end{prop}
$\quad$\\
$\bullet$ \underline{Computation of the real kernel of the Hessian}\\\\
If $A\in M_n(\cc)$, then we define $\ker_{_{\rr}}(A):=\{ x\in\rr^n : A(x)=0 \}=\ker(\Re(A))\cap\ker(\Im(A))$.\\
\begin{prop}\label{calculgnoyau}
Let $(t,z)\in \ceg$. Then the real kernel of the Hessian is : $\ker_{_{\rr}}\mbox{Hess} \;\vfi_{E,g}(t,z)=$
\begin{equation}\label{noyau}
\{(\tau,\aaa)\in \rr\times \rr^{2d}: \aaa \perp \nabla 
H (z), \tau J\nabla H(z)+(\mg F_z(t)-Id)\aaa =0 \}.
\end{equation}
\end{prop}
$\quad$\\
\underline{{\it Proof}} : Let $\tau\in\rr$ and $\aaa\in\rdd$. We set:
$$x:=\tau \jgrad +(\mg F_z(t)-I)\aaa.$$
Let us denote by $\wtu$ and $\wtd$ the real and imaginary part of $\wt$. Then,
$(\tau,\aaa)\in\ker_{_{\rr}}\mbox{Hess} \;\vfi_{E,g}(t,z)$ if and only if:
\beq\label{heartsuit}
<\wtd \jgrad;x>=2<\grad;\aaa>.
\eeq
\beq\label{diamondsuit}
<(I-\wtu) \jgrad;x>=0.
\eeq
\beq\label{clubsuit}
(^tF_z(t)^t\mg-I)(I-\wtu)x =0.
\eeq
and
$$-2\tau \grad  + [J\mg F_z(t)-^t(\mg F_z(t))J]\aaa + (^tF_z(t)^t\mg-I)\wtd x=0.$$
We multiply this last identity by $(\mg F_z(t))J$, we note that $\wtd=J\wtu$ and recall that $\mg F_z(t)$ is
symplectic to obtain the equivalent identity:
\beq\label{spadesuit}
(\mg F_z(t)-I)(\wtu-I)x=2x.
\eeq
Now, if $(\tau,\aaa)\in\ker_{_{\rr}}\mbox{Hess} \;\vfi_{E,g}(t,z)$, then, by (\ref{spadesuit}) and
(\ref{clubsuit}), we have:
$$<x,(I-\wtu)x>=0,\mbox{ i.e. } |x|^2=<\wtu x,x>.$$
By lemma \ref{normWt}, $\norm{\wtu}_{\mathcal{L}(\rr^{2d})}<1$, thus $x=0$, and by (\ref{heartsuit}),
$\grad\perp\aaa$.\

Conversely, if $x=0$ and $\grad\perp\aaa$, then, we have (\ref{heartsuit}), (\ref{diamondsuit}), 
(\ref{clubsuit}) and (\ref{spadesuit}). Thus $(\tau,\aaa)\in \ker_{_{\rr}}\mbox{Hess}
\;\vfi_{E,g}(t,z)$.\qed\\\

We are now able to describe the conditions under which we can apply the generalised stationary phase
theorem on $L_{\eta,\nu,\mu,\gmm,j}(h)$: we easily check the positivity of the imaginary part of the phase
$\vfi_{E,g}$ by lemma \ref{normWt}. Moreover, if $\ceg$ is a union of smooth submanifolds of
$\rr\times\rdd$, if $X\in\ceg$, then the Hessian of $\vfi_{E,g}(X)$ is non-degenerate on the normal space
$N_X\ceg$ if and only if $\ker_{_{\rr}}\mbox{Hess} \;\vfi_{E,g}(X)\subset T_X\ceg$, the tangent space of
$\ceg$ at $X$. Besides, note that, by the non-stationary phase theorem, we can restrict this hypothesis
to points $X$ in $\supp(\hat{f})\times \supp(a_{j_0})$.\\\\
{\it Definition}: let $g\in G$, $T>0$, such that $\supp(\hat{f})\subset]-T,T[$, and $\Psi_g:=\left\{
\barr{l}
]-T,T[\times \nrj\to \rdd\\
(t,z)\mapsto \mg\Phi_t(z)-z
\earr\right.$
We say that `the flow is $g$-clean on $]-T,T[\times \nrj$' if zero is a weakly regular value of
$\Psi$, i.e. :
\begin{itemize}
\item $\Psi_g^{-1}(\{ 0 \})=:\ceg$ is a finite union of smooth submanifolds of $\rr\times\rdd$.
\item $\forall (t,z)\in \ceg$, $\quad T_{(t,z)}\ceg=\ker d_{(t,z)}\Psi_g$.
\end{itemize}
We say that `the flow is $G$-clean on $]-T,T[\times\nrj$' if it is $g$-clean for all $g$ in $G$.\\\

By proposition \ref{calculgnoyau}, we see that if $(t,z)\in\ceg$, then $\ker
d_{(t,z)}\Psi_g=\ker_{_{\rr}}\mbox{Hess} \;\vfi_{E,g}(t,z)$. Thus, if we only know that the support of
$\hat{f}$ is in $]-T,T[$, then the $g$-clean flow condition is the minimal hypothesis under which we can apply the
stationary phase theorem to $L_{\eta,\nu,\mu,\gmm,j}(h)$. Therefore, we can state the theorem:
\begin{theo}\label{flotpropre} {\bf Reduced trace formula with $G$-clean flow.}\\
Let $G$ be a finite subgroup of $Gl(\rr,d)$ and $H:\rdd\to\rr$ a smooth Hamiltonian $G$-invariant.
Suppose that $E\in\rr$ is such that there exists $\delta E>0$ such that $H^{-1}([E-\delta E, E+\delta
E])$ is compact, and $\nrj=\{ H=E\}$ has no critical points. Make hypothesis (\ref{hypCF}). Let $f$ and
$\psi$ be real functions in
$\mathcal{S}(\rr)$ such that $\supp(\psi)\subset ]E-\delta E,E+\delta E[$ and $\hat{f}$ is compactly
supported in $]-T,T[$, where $T>0$. Suppose that the flow is $G$-clean on $]-T,T[\times\nrj$. Then the spectral density
$$\mathcal{G}_{\chi}(h)=\frac{d_{\chi}}{|G|}\sum_{g\in G}\overline{ \chi(g) } \, I_{g,E}(h)  -see\;
(\ref{densitespectrale}),\; (\ref{Ig})-$$
has a complete asymptotic expansion as $h\to 0^+$. Moreover, if $g\in G$, and, if $[\ceg]$ denotes the set of connected
components of $\ceg$, then the quantity $\int_0^t p_s \dot{q_s}ds$ is constant on each element $Y$ of
 $[\ceg]$, denoted by $S_{Y,g}$, and we have the following expansion:
$$I_{g,E}(h)=\sum_{Y\in [\ceg]} (2\pi h)^{\frac{1-\dim Y}{2}} e^{\frac{i}{h}S_{Y,g}} \frac{1}{2\pi}
\left( \int_Y \hat{f}(t)\,\psi(E) d_g(t,z) d\sigma_Y(t,z) +\sum_{j\geq 1}h^j a_{j,Y}\right)+O(h^{+\infty})$$
where $a_{j,Y}$ are distributions in $\hat{f}\otimes(\psi\circ H)$ with support in $Y$, and the density $d_g(t,z)$ is defined by:
\beq\label{density}
d_g(t,z):=\detp\left( \frac{\varphi_{E,g}^{\prime\prime}(t,z)_{|_{\mathcal{N}_{(t,z)}Y}}}{i} \right)
\detp\left( \frac{A+iB-i(C+iD)}{2}\right).
\eeq
$\varphi_{E,g}$ is given by (\ref{phase1}) and $A, B, C, D$ are the $d\times d$ blocs forming the
matrix $F_z(t):=\partial_z(\Phi_t(z))$ (see (\ref{monodr})).
\end{theo}
\textsl{Remark}: without symmetry, this theorem can be compared to articles of T.Paul and A.Uribe (cf
\cite{P-U} and \cite{P-U2}) or to the Gutzwiller formula in the PhD. thesis of S.Dozias (\cite{Do}), see also
\cite{Mein}.
A notion of clean flow is also present in \cite{CRR}. The density $d_g(t,z)$ is difficult to compute in
general, even without symmetry. The purpose of next sections is to calculate it in two special cases: when
$\hat{f}$ is supported near zero (Weyl part), and under an assumption of non-degenerate periodic orbits
of the classical flow in $\nrj$ (oscillating or Gutzwiller part).\\\\
\underline{{\it Proof}} : as we have seen before, we can apply the stationary phase theorem on each
$L_{\eta,\nu,\mu,\gmm,j}(h)$, which gives an expansion of each $I_g^{j_0}(h)$ and each $I_g(h)$. The first
term is given by:
$$I_g(h)\underset{h\to 0^+}{\sim} \frac{(2\pi h)^{-d}}{2\pi}\int_{\rr_t}\gra \chi_2(\aaa) \hat{f}(t)
\psi(H(\aaa)) \detp\left( \frac{A+iB-i(C+iD)}{2}\right) e^{\frac{i}{h}\varphi_{E,g}(t,\aaa)}dtd\aaa.$$
By definition of $\ceg$, $\varphi_{E,g}$ is constant on each connected component of $\ceg$, equal to:
$$\varphi_{E,g}(t,\aaa)=S(\aaa,t)+Et=\int_0^tp_s \dot{q_s}ds, \quad\mbox{ where } (q_s,p_s)=\Phi_s(\aaa).$$
This ends the proof of theorem \ref{flotpropre}.\qed
\subsection{The Weyl part}\label{Weylpart}
We now deal with one case which leads to an asymptotic expansion at the first order of the counting function of $\hqr$ in an interval of
$\rr$. Fix $g$ in $G$ and define:
\beq
\mathcal{L}_{E,g}:=\{ t\in\rr:\;\exists z\in\nrj:\;\mg\Phi_t(z)=z \}.
\eeq
\begin{theo}\label{Ig_W}
Let $G$ be a finite subgroup of $Gl(\rr,d)$ and $H:\rdd\to\rr$ a smooth $G$-invariant Hamiltonian.
Let $E\in\rr$ be such that $H^{-1}([E-\delta E, E+\delta E])$ is compact for some $\delta E>0$, and that
$\nrj=\{ H=E\}$ has no critical points. Make hypothesis (\ref{hypCF}). Let $f$ and $\psi$ be real functions
in $\mathcal{S}(\rr)$ with $\supp(\psi)\subset ]E-\delta E,E+\delta E[$ and $\hat{f}$ is compactly
supported.
For $g$ in $G$, we set:
$$\nu_g:=\dim \ker (g-Id_{\rr^d}), \quad F_g:=\ker (M(g)-Id_{\rr^{2d}}) \quad\mbox{ and }\quad
\tilde{F}_g:=\ker (g-Id_{\rr^{d}}).$$
Set $I_g(h):=Tr\left(\psi(\hq) f\left( \frac{E-\hq}{h} \right)\mm\right)$.
Then, under previous assumptions, we have:\\
-- If $\supp \hat{f}\cap \mathcal{L}_{E,g}=\emptyset$, then
$I_{g,E}(h)=O(h^{+\infty})$.\\
-- If $\supp \hat{f}\cap \mathcal{L}_{E,g}=\{ 0 \}$ then
 we have the following expansion modulo
$O(h^{+\infty})$ :
\begin{equation}\label{weylasymp}
I_{g,\la}(h) \asymp h^{1-\nu_g}\sum_{k\geq 0} c_k(\hat{f},g)h^{k}, \quad \mbox{ as } h\to 0^+.
\end{equation}
uniformly in $\lambda$ in a small neighborhood of $E$, where $c_k(\hat{f},g)$ are distributions in
$\hat{f}$ with support in $\{ 0 \}$, and, if  $d(\nrjl\cap F_g)$ denotes the euclidian measure on
$\nrjl\cap F_g$, then we have:
\begin{equation}\label{Weyl}
c_0(\hat{f},g)= \psi(\la)\hat{f}(0)\frac{(2\pi)^{-\nu_g}}{\det ((Id_{\rr^d}-g)|_{\tilde{F}_g^{\perp}})}
\int_{\nrjl\cap F_g}\frac{d(\nrjl\cap F_g) (z)}{|\nabla H(z)|}.
\end{equation}
\end{theo}
\textsl{Remark 1}: the oscillating term of Theorem \ref{flotpropre} is now vanishing, since, for $g\in G$,
$S_{Y,g}=0$ when $Y=\{ 0 \}\times (\nrj\cap F_g)$. Moreover, it is easy to see that, since $\nrj$ is compact and
non-critical, zero is isolated in $\mathcal{L}_{E,g}$. Thus the hypothesis $\supp \hat{f}\cap \mathcal{L}_{E,g}=\{ 0 \}$
is fulfilled if $\hat{f}$ is supported close enough to zero.

\textsl{Remark 2}: we slightly precised the previous result of Z. El Houakmi given in \cite{El.H}, by
the computation of (\ref{Weyl}). Note that the leading term of  $\mathcal{G}_\chi(h)$ is obtained for $g=Id$,
and:
$$\mathcal{G}_\chi(h)=\frac{d_\chi^2}{|G|}\psi(E)\hat{f}(0)
(2\pi h)^{1-d}\frac{1}{2\pi}\int_{\nrj}\frac{d\nrj}{|\nabla H|} + O(h^{2-d}), \;\mbox{ as } h\to 0^+.$$
\underline{ {\it Proof}} : If $\supp \hat{f}\cap \mathcal{L}_{E,g}=\emptyset$, then
$(\supp(\hat{f})\times\rdd)\cap\ceg=\emptyset$, and by the non stationary phase theorem, we get the result.\\
Now suppose that $\supp \hat{f}\cap \mathcal{L}_{E,g}=\{ 0 \}$. Then we have:
\beq
\ceg\cap(\supp(\hat{f})\times\rdd)=\{ 0 \}\times (\nrj\cap F_g).
\eeq
\underline{{\it We now give some `trick' to boil down to the case where $G$ is composed of isometries}}. We recall that, since $G$ is compact,
there is some $S_0$, symmetric $d\times d$ positive definite matrix, such that:
\beq\label{feintajul}
G_0:=S_0^{-1} \, G \, S_0 \mbox{ is a subgroup of the orthogonal group } O(d,\rr).
\eeq
One can indeed classicaly find a scalar product invariant by $G$ by averaging with the Haar measure of $G$.
Thus, we can define a new $G_0$-invariant Hamiltonian:
$$H_0(z):=H(M(S_0)z),\qquad where \quad M(S_0):=\left(
\barr{cc}
S_0 & 0\\
0   & ^tS_0^{-1}\\
\earr\right).$$
If $\chi\in \widehat{G}$, then one can define $\chi_0:G_0\to \cc$ by:
$$\chi_0(g_0):=\chi(S_0 \, g_0\, S_0^{-1}).$$
Then it is easy to check that $\chi_0\in \widehat{G}_0$ and that the application $\chi\mapsto\chi_0$ is
bijective from $\widehat{G}$ to $\widehat{G}_0$. Moreover, identity (\ref{metaplec}) implies that:
$$\op(H_0)=\tilde{M}(S_0)^{-1}\op(H) \tilde{M}(S_0).$$
If $\chi\in \widehat{G}$, then we can define:
$$\tilde{P}_{\chi_0}:=\frac{d_{\chi_0}}{|G_0|}\sum_{g_0\in G_0}\overline{\chi_0(g_0)} \, \tilde{M}(g_0).$$
Then we have
$\tilde{P}_{\chi_0}=\tilde{M}(S_0)^{-1} P_\chi \tilde{M}(S_0)$. Therefore, if $f(\hq)$ is trace class, then
$f(\hq_0)$ also, and we have:
$$\tr(f(\hqr))=\tr(f(\hq) P_{\chi})=\tr(f(\hq_0)\tilde{P}_{\chi_0})),$$
by cyclicity of trace. This remark apply in particular for the trace (\ref{densitespectrale}). Moreover, if
$g\in G$, if $g_0:=S_0^{-1} g S_0$, then $\tr(f(\hq)\mm)=\tr(f(\hq_0)\tilde{M}(g_0))$.
Finally, it is easy to check that hypotheses for ($H$, $G$) are available for ($H_0$, $G_0$), and that
coefficients of the asymptotic have the same expression in terms of ($H_0$, $G_0$) as in ($H$, $G$).\qed\\\

From now on, we suppose that $G$ is made of isometries, without loss of generality.\\
First, we remark that $\nrj$ and $F_g$ are transverse submanifolds of $\rdd$. Indeed, if $z\in\nrj\cap F_g$,
then, by (\ref{grad}), since $g$ is an isometry, we have $\grad\in F_g$, thus $F_g+[\rr\grad]^\perp=\rdd$.
Therefore
$$\mathcal{T}_{(0,z)}\ceg=\{ 0 \}\times [F_g\cap [\rr\nabla H(z)]^{\perp}].$$
If $(\tau,\aaa)\in \ker_{_{\rr}}\mbox{Hess} \; \vfi_E(0,z)$ then by Proposition \ref{calculgnoyau},
$\tau J\nabla H(z)+(\mg -I_{2d})\aaa =0 $. Then one can take the scalar product of this equality with
$\jgrad$ to obtain $\tau=0$ and thus, $\ker_{\rr}\mbox{Hess} \; \vfi_E(0,z)=\mathcal{T}_{(0,z)}\ceg$. This
means that we have the theorical asymptotic expansion of Theorem \ref{Ig_W}.\\\

Now, we have to compute the leading term of this expansion. Here again, we can suppose that $g$ is an
isometry, which simplifies the calculus: in particular, $[\mg,J]=0$, when $t=0$, we have $\wt=0$, and
$F_z(0)=Id$. By Proposition \ref{calculhess}, we obtain:
$${\small \mbox{Hess} \; \vfi_{E,g}(0,z)=\left(
\barr{c|c}
\frac{i}{2}|\nabla H(z)|^2 & -^t\nabla H(z)\\\hline

-\nabla H(z) & \ud J(M(g)-M(\gmu))+\frac{i}{2}(I-M(g))(I-M(\gmu)) \\
\earr
\right).}$$
We have $\mathcal{N}_{(0,z)} \ceg=\rr\times [F_g^{\perp}+\rr\grad]$. Let $\beta_0$ be a basis of
$F_g^{\perp}$.  We set:
$$e_0:=\frac{\partial }{\partial t}=(1,0), \qquad \e_0:=(0,\grad).$$
Let $\beta$ be the basis of $\mathcal{N}_{(0,z)} \ceg$ made up of (in this order) $e_0$, $\e_0$ and $\beta_0$. We
note that the linear application $\ud J(\mg-\mgm)+\frac{i}{2}(I-\mg)(I-\mgm)$ stabilizes the space $F_g^{\perp}$. Then by calculating
the determinant of the restriction of $\mbox{Hess} \; \vfi_E(0,z)$ to $\mathcal{N}_{(0,z)} \ceg$ in this
basis, we get (noting $\mathcal{N}:=\mathcal{N}_{(0,z)}\ceg$):
$$\det\left(\frac{\varphi_{E,g}^{\prime\prime}(0,z)_{|_{\mathcal{N}}}}{i}\right)=
|\nabla H(z)|^2 \det \left[\frac{1}{2i}J(\mg-\mgm)+\frac{1}{2}(I-\mg)(I-\mgm)\right]_{|_{F_g^{\perp}}}$$
If $\Pi_g$ is the orthogonal projector on $\widetilde{F}_g$, then we have:
$$\frac{1}{|\grad|^{2}}\det\left(\frac{\varphi_{E,g}^{\prime\prime}\mbox{{\scriptsize $(0,z)$}}_{|_{\mathcal{N}}}}{i}\right)=
\left(
\barr{c|c}
\frac{1}{2}(I_d-g)(I_d-\gmu) + \Pi_g &  \frac{1}{2i}(g-\gmu)\\\hline

 -  \frac{1}{2i}(g-\gmu)                         & \frac{1}{2}(I_d-g)(I_d-\gmu) + \Pi_g\\
\earr
\right).$$
Then, since $g$ is an isometry, we can suppose that $g$ is bloc diagonal with blocs $I_{p_1}$, $-I_{p_2}$,
$R_{\theta_1}, \dots, R_{\theta_r}$, where $p_1+p_2+2r=d$, $\theta_j$'s are not in $\pi \zz$, and
$R_{\theta}:=\left(
\begin{array}{cc}
\cos \theta    & -\sin \theta \\
\sin \theta    & \cos \theta  \\
\end{array}
\right).$
We then use the fact that $g$ commutes with $\Pi_g$, and that when $[C,D]=0$, then $\det${\scriptsize $\left(
\barr{cc}
A & B\\
C & D
\earr\right)$}
$=\det(AD-BC)$, for any blocs $A$, $B$, $C$, $D$ of same size. A straightforward calculus then gives (see
\cite{these} for details):
$$\det\left(\frac{\varphi_{E,g}^{\prime\prime}(0,z)_{|_{\mathcal{N}_{(0,z)}\ceg}}}{i}\right)=
|\grad|^{2}\det  \left[(I_d-g)_{|_{\widetilde{F}_g^{\perp}}}\right]^2.$$
Since $\left[\detp\right]^2=\det$, we have:
$$\detp\left(\frac{\varphi_{E,g}^{\prime\prime}(0,z)_{|_{\mathcal{N}_{(0,z)}\ceg}}}{i}\right)=
\pm |\grad|| \det(I_d-g)_{|_{\widetilde{F}_g^{\perp}}}|.$$
We can proove that the factor $\pm 1$ is in fact equal to $1$, either by coming back to the calculus of
$\detp$ with gaussians, or, classically, by using a weak asymptotic, i.e. by calculating the asymptotic of
$\tr(\vfi(\hq)\mg)$, when $\vfi:\rr\to\rr$ is smooth and $\vfi(\hq)$ is trace class. See \cite{these} for
details.\

Using (\ref{density}), the fact that the phase vanishes on $\ceg$, and that
$\dim(\ceg)\cap(\supp(\hat{f})\times\rdd)=2\nu_g-1$, we obtain the result we claimed.
This ends the proof of Theorem \ref{Ig_W}.\qed\\\\
As a consequence of Theorem \ref{Ig_W} near $t=0$, using a well known Tauberian argument (see \cite{Ro}),
we get the following:
\begin{cor}\label{comptage}
Let $G$ be a finite group of $Gl(d,\rr)$, $H:\rdd\to\rr$ a $G$-invariant smooth Hamiltonian satisfying
(\ref{hypCF}). Let $E_1<E_2$ in $\rr$, and $I:=[E_1,E_2]$. Suppose that there exists $\varepsilon >0$ such that
$H^{-1}([E_1-\varepsilon,E_2+\varepsilon])$ is compact.
Furthermore suppose that $E_1$ and $E_2$ are not critical values of $H$.
If $\chi\in\widehat{G}$, then the spectrum of $\hq_{\chi}$ is discrete in $I$, and we have:
$$N_{I,\chi}(h)=\frac{d_{\chi}^2}{|G|}
(2\pi h)^{-d}\mbox{Vol}\,[H^{-1}(I)]+O(h^{1-d}),$$
where $N_{I,\chi}(h)$ is the number of eigenvalues of $\hqr$ in $I$ counted with multiplicity.
\end{cor}
\textsl{Remark}: One can interpret this result by saying that, semi-classically, the proportion of
eigenfunctions of $\hq$ having symmetry $\chi$ is $\frac{d_{\chi}^2}{|G|}$. In particular, the same
proportion of eigenvalues has multiplicity greater than $d_\chi$. The more $d_\chi$ is high, the more $\ldec$
takes part in the spectrum of $\hq$.
\subsection{The oscillatory part}\label{Gutzpart}
\begin{figure}[h!]
   \epsfig{file=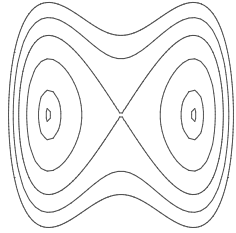,width=3cm,height=3cm}
   \caption{Double well phase portrait}
   \label{traj}
\end{figure}

If $g\in G$ and $\gmm$ is a periodic orbit of $\nrj$ globally stable by $M(g)$, we set :
$$\mathcal{L}_{g,\gmm} :=\{ t\in \supp \hat{f} : \exists z\in \gmm :\mg\Phi_t(z)=z \}.$$
If $t_0\in \mathcal{L}_{g,\gmm}$, $z\in\gmm$, then  $P_{\gmm,g,t_0}$ denotes the Poincar\'e map
of $\gmm$ between $z$ and $\mgm z$ at time $t_0$, restricted to $\nrj$. The characteristic polynomial of
$dP_{\gmm,g,t_0}$ doesn't depend on $z\in\gmm$. Note that, by iterating formula (\ref{flowsym}), since $G$ is finite,
if we have $\mg\Phi_t(z)=z$, then $z$ is a periodic point of the Hamiltonian system (\ref{hamsyst}). 
\begin{theo}\label{Gutz_osc}
Make the same assumptions as in Theorem \ref{Ig_W}, but suppose that $0\notin \supp \hat{f}$.
Make the following hypothesis of non-degeneracy : if $\gmm\subset\nrj$, is such that $\exists g\in G$ and
$\exists t_0\in \mathcal{L}_{g,\gmm}$, $t_0\neq 0$, then $1$ is not an eigenvalue of $\mg dP_{\gmm,g,t_0}$.
Then the set of such $\gmm$'s is finite and the following expansion holds true modulo $O(h^{+\infty})$, as $h\to 0^+$ :
$$\mathcal{G}_{\chi}(h)\asymp \frac{d_{\chi}}{|G|}
\sum_{\tiny{\barr{c}
\gmm  \mbox{ periodic }\\
\mbox{orbit of } \nrj
\earr}}
\sum_{\tiny{\barr{c}
g\in G  \mbox{ s.t. }\\
M(g)\gmm = \gmm
\earr}}
 \overline{ \chi(g) }
\sum_{\tiny{\barr{c}
t_0\in\mathcal{L}_{g,\gmm}\\
t_0\neq 0
\earr}}
e^{\frac{i}{h} S_{\gmm}(t_0)} \sum_{k\geq 0}d_k^{\gmm ,g,t_0}(\hat{f})h^k.$$
Terms $d_k^{\gmm,g,t_0}(\hat{f})$ are distributions in $\hat{f}$ with support in $\{ t_0 \}$,
$ S_{\gmm}(t_0):=\int_0^{t_0}p_s \dot{q}_s ds $,  $((q_s,p_s):=\Phi_s(z) \mbox{ with }z\in\gmm)$,
and 
$$d_0^{\gmm ,g,t_0}(\hat{f})=\frac{\psi(E)\, T_{\gmm}^*\,  e^{i \frac{\pi}{2} \sigma_{\gmm}(g,t_0)}}
{2\pi |\det(\mg dP_{\gmm,g,t_0}-Id)|^{\frac{1}{2}}} \hat{f} (t_0)$$
where $T_{\gmm}^*$ is the primitive period of $\gmm$ and $\sigma_{\gmm}(g,t_0)\in \zz$.
\end{theo}
{\it Example 1}: if $d=1$, periodic orbits are always non-degenerate. For example, in the case of a
double well Schr\"odinger Hamiltonian, one can illustrate the sum of Theorem \ref{Gutz_osc} on figure
\ref{traj}, picturing the classical flow in $\rr^2$: some periodic orbits appear only for $g=Id$ in the
sum, and others arise for both $g=\pm Id$. One can also fold the picture to compare with the periodic
orbits of the reduced space as in Theorem \ref{Gutzreduit}.

{\it Example 2}: if $H$ is a Schr\"odinger operator on $\rd$ with potential $V(x)=<Sx,x>$, where $S$ is
the diagonal matrix with diagonal non-vanishing $w_1^2,\dots,w_d^2$, if one assumes that $\forall
i\neq j$, $w_i/w_j\notin \mathbb{Q}$, then periodic orbits appear as a union of $d$ plans, with primitive
periods $T_j^*=\frac{\pi}{w_j}$ and are all non-degenerate.\\\

As a particular case of this theorem, we get the Theorem \ref{Gutzreduit}:\\\\
\underline{{\it Proof of Theorem \ref{Gutzreduit}}}: if we suppose that $G$ acts freely on $\nrj$, then $\nrj/G$
inherits a structure of smooth manifold such that the canonical projection $\pi:\nrj\twoheadrightarrow\nrj/G$ is smooth, and the
dynamical system restricted to $\nrj$ descends to quotient. If $t_0\in\rr^*$, $g\in G$ and $z\in\nrj$,
with orbit $\gmm$, are such that $\mg\Phi_{t_0}(z)=z$, then $\gmm$ and $\pi(\gmm)$ are periodic.
If $P_{\pi(\gmm),\pi(z)}(t_0)$ denotes the Poincar\'e map of $\pi(\gmm)$ at time $t_0$, then we have:
\beq\label{Poincarered}
\det(\mg d_z P_{\gmm,g,t_0}-Id)=\det(d_{\pi(z)} P_{\pi(\gmm),\pi(z)}(t_0)-Id).
\eeq
Indeed, if $\tilde{\Phi}_t$ denotes the flow in $\nrj/G$, then one can differentiate the following
identity on $\nrj$ with variable $z$:
$$\pi(\mg \Phi_{t_0}(z))=\tilde{\Phi}_{t_0}(\pi(z)),$$
to get the the identity:
$$d_z\pi \circ \mg F_z(t_0)= \tilde{F}_{\pi(z)}(t_0)\circ d_z\pi,$$
where $\tilde{F}_{\pi(z)}(t_0)$ is the differential of $x\mapsto \tilde{\Phi}_{t_0}(x)$ at $\pi(z)$.
Moreover, $\pi$ is a submersion, and by a dimensional argument it's also an immersion. Thus we have
(\ref{Poincarered}).\

Therefore, if we make hypotheses of Theorem \ref{Gutzreduit}, then hypotheses of Theorem \ref{Gutz_osc} are
fulfilled. If $z\in\nrj$ is such that the orbit of $\pi(z)$ is periodic with period $t_0\neq 0$, then there
is only one $g=g_\gmm\in G$ such that $\mg \Phi_{t_0}(z)=z$. If $\mathcal{L}_{red}$ denotes the set of
periods of $\nrj/G$, then we have:
$$\sum_{\tiny{\barr{c}
\gmm  \mbox{ periodic }\\
\mbox{orbit of } \nrj
\earr}}
\sum_{\tiny{\barr{c}
g\in G  \mbox{ s.t. }\\
M(g)\gmm = \gmm
\earr}}
\sum_{\tiny{\barr{c}
t_0\in\mathcal{L}_{g,\gmm}\\
t_0\neq 0
\earr}}\dots
=\sum_{t_0\in\mathcal{L}_{red}} 
\sum_{\substack{\gmm \subset \nrj:\pi(\gmm)\mbox{ {\tiny periodic}}\\
\mbox{ {\tiny with $t_0$ for period }}}}\sum_{g=g_\gmm}\dots.$$
If we denote $Stab(\gmm):=\{ g\in G : \mg \gmm=\gmm\}$, then we have $Stab(\gmm)=<g_\gmm>$ and it is easy to
see that $T_{\pi(\gmm)}^*=\frac{T_{\gmm}^*}{|Stab(\gmm)|}$. If we denote by $N_{\pi(\gmm)}$ the number of
orbits of $\nrj$ with image $\pi(\gmm)$ by $\pi$, then we have $N_{\pi(\gmm)}=|G|/|Stab(\gmm)|$.
Thus we have:
$$\mathcal{G}_{\chi}(h)=d_{\chi} 
\sum_{t_0\in\mathcal{L}_{red}} \hat{f} (t_0)
\sum_{\substack{\gmm \subset \nrj:\pi(\gmm)\mbox{ {\tiny periodic}}\\ \mbox{ {\tiny with $t_0$ for period }}}}
\overline{ \chi(g_{\pi(\gmm)}(t_0)) } \frac{T_{\pi(\gmm)}^*}{N_{\pi(\gmm)}} 
\frac{e^{\frac{i}{h} S_{\gmm}(t_0) } e^{i \frac{\pi}{2} \sigma_{\gmm}(g,t_0)}}
{2\pi |\det(d_{\pi(z)} P_{\pi(\gmm),\pi(z)}(t_0)-Id)|^{\frac{1}{2}}}
 + O(h).$$
Then one can show that quantities appearing in the r.h.s. don't depend on $\gmm$ but only on $\pi(\gmm)$, and
this proves the Theorem \ref{Gutzreduit}.\qed\\\\
\underline{{\it Proof of the Theorem \ref{Gutz_osc}}}:
We fix $g$ in $G$. If $t_0\in\rr^*$, we set:
$$\Gamma_{E,g,t_0} :=\{\gmm \mbox{ orbit of } \nrj \;:\; \exists z\in\gmm\; :\; \mg \Phi_{t_0}(z)=z \}.$$
\begin{lem}
If we make assumptions of non-degeneracy of Theorem \ref{Gutz_osc}, then $\mathcal{L}_{E,g}\cap \supp(\hat{f})$ is
finite and we have:
\beq
\ceg\cap(\supp(\hat{f})\times\rdd)=\bigcup_{\substack{t_0\in\mathcal{L}_{E,g}\\ t_0\neq 0} }\bigcup_{\gmm \in
\Gamma_{E,g,t_0}} \{ t_0 \}\times \gmm.
\eeq
\end{lem}
\underline{{\it Proof}} : one can adapt the proof of the cylinder theorem of \cite{Ab-Ma}. For details, we
refer to \cite{these}.\qed\\\

Note that periodic orbits appearing in this critical set are the ones stable by $g$.\\
We see that $\ceg\cap(\supp(\hat{f})\times\rdd)$ is a submanifold of $\rr\times\rdd$ and if $(t_0,z)\in\ceg$,
then we have:
$$T_{(t_0,z)}\ceg=\{ 0 \}\times \rr\jgrad.$$
To apply the stationary phase theorem, we have to show that $\ker_{_{\rr}} \mbox{Hess}\;
\varphi_{E,g}(t_0,z)\subset T_{(t_0,z)}\ceg$. Let $(\tau,\aaa)\in\ker_{\rr} \mbox{Hess}\;
\varphi_{E,g}(t_0,z)$. By Proposition \ref{calculgnoyau}, we have $\aaa\perp \grad$ and:
\beq\label{wizz}
\tau \jgrad + (\mg F_z(t_0)-I)\aaa=0.
\eeq
If $\la\in\rr$, we denote by $E_{\lambda}:=\sum_{k=1}^{2d}\ker(\mg F_z(t_0)-Id)^k$. Let $\gmm$ be the orbit of $z$.
Since $1$ is not an eigenvalue of $\mg dP_{\gmm, g,t_0}$, $1$ is an eigenvalue of $\mg F_z(t_0)$ of
multiplicity $2$. Thus $\dim E_1=2$. Using (\ref{vecpropre}) and (\ref{wizz}), we have $\aaa\in E_1$. Let
$u_2\in\rdd$ such that $(\jgrad,u_2)$ is a basis of $E_1$.  Note that $<u_2,\grad>\neq 0$, otherwise we
would have $u_2\in (JE_1)^{\perp}$, which is equal to $\underset{\la\neq 1}{\oplus} E_{\la}$ since
$\mg F_z(t_0)$ is symplectic. Since $\aaa\in E_1$ we have $\la_1,\la_2$ in $\rr$ such that:
$$\aaa=\la_1 \jgrad + \la_2 u_2.$$
Then, using the fact that $<\aaa,\grad>=0$, we get $\la_2=0$ (since  $<u_2,\grad>\neq 0$). Thus coming back
to (\ref{wizz}), we get $\tau=0$ and $\aaa\in\rr\jgrad$. Thus $(\tau,\aaa)\in T_{(t_0,z)}\ceg$.\\\

This shows that we can apply the stationary phase theorem and get a  theorical expansion of $I_g(h)$ and
$\mathcal{G}_\chi(h)$. We have now to compute the first term of this expansion.
We suppose that $(t_0,z)\in\ceg$. We denote by $\Pi$ the orthogonal projector on $\rr\jgrad$. We set
$F:=\mg F_z(t_0)$ and $W:=\widehat{W}_{t_0}$. Then we have:
$\det\left( \frac{\varphi_{E,g}^{\prime\prime}(t_0,z)_{|_{\mathcal{N}_{(t_0,z)}\ceg}}}{i} \right)=$
$$\det\left(
\barr{c|c}
\frac{1}{2}<(I-W) \jgrad;\jgrad> & -\frac{1}{i} ^t\grad        \\
 & + \frac{1}{2}^t\left[(^tF-I)(I-W) \jgrad \right] \\\hline

-\frac{1}{i}\grad & \frac{1}{2i}[JF +^t(JF)] \\
+\frac{1}{2}(^tF-I)(I-W) \jgrad & +\frac{1}{2}(^tF-I)(I-W)(F-I)+\Pi\\
\earr
\right).$$
Since $F$ is symplectic, we have $JF+^t(JF)=(^tF+I)J(F-I).$
Set:
\beq
K:=\frac{1}{2i} (^tF+I)J+\ud (^tF-I)(I-W).
\eeq
Then, the forth bloc is equal to $K(F-I)+\Pi$.\\
Using (\ref{vecpropre}), we note that the third bloc is equal to $K\jgrad$.
Let us set:
\beq
X_1:=\ud (I-W)\jgrad.
\eeq
We then have:
$$\det\left( \frac{\varphi_{E,g}^{\prime\prime}(t_0,z)_{|_{\mathcal{N}_{(t_0,z)}\ceg}}}{i} \right)=
\det\left(
\barr{c|c}
^tX_1\jgrad & i^t\grad +^tX_1(F-I)\\\hline
K\jgrad     & K(F-I) +\Pi
\earr\right)$$
The following technical lemma is due to M. Combescure (see \cite{CRR} in the preprint version or \cite{these}
 p.87 for the proof):
\begin{lem}\label{combescuretechnic}
$K$ is invertible and $K^{-1}=\ud [(F-I)+i(F+I)J]$.\\
Moreover, if we set $F=\left(
\barr{cc}
\tilde{A} & \tilde{B} \\
\tilde{C} & \tilde{D}
\earr\right)$, then $\det(K)=(-1)^d
\det(\ud(\tilde{A}+i\tilde{B}-i(\tilde{C}+i\tilde{D})))^{-1}.$

\end{lem}
Since
$$\det\left( \frac{\varphi_{E,g}^{\prime\prime}(t_0,z)_{|_{\mathcal{N}_{(t_0,z)}Y}}}{i} \right)=
\det\left(
\barr{c|c}
1 & 0\\\hline
0 & K
\earr
\right)
\left(\barr{c|c}
^tX_1\jgrad & i^t\grad +^tX_1(F-I)\\\hline
\jgrad     & (F-I) +K^{-1}\Pi
\earr
\right),$$
using (\ref{density}) and the preceeding lemma, we get:
\beq\label{densite}
d_g(t,z)^{-2}=(-1)^d\det(\gmu)\det\left(\barr{c|c}
^tX_1\jgrad & i^t\grad +^tX_1(F-I)\\\hline
\jgrad     & (F-I) +K^{-1}\Pi
\earr
\right).
\eeq
We denote by $\aaa:=<X_1,\jgrad>$\footnote{NB : $\aaa\neq 0$ since $I-W$ is invertible and $\jgrad\neq 0$.}
and we use the line operation $L_2\gets L_2-\frac{1}{\aaa} \jgrad L_1$, to get:
\beq\label{densite2}
d_g(t,z)^{-2}=(-1)^d \aaa \det(D)\det(\gmu).
\eeq
where
$$D:=(F-I) +K^{-1}\Pi-\frac{1}{\aaa}\jgrad [i^t\grad +^tX_1(F-I)].$$
Then, we compute $\det(D)$ in the basis $\beta_0:=(v_1,\dots,v_{2d})$ where
$v_1:=\jgrad$, $v_2$ is such that $v_2\perp \jgrad$ and $(v_1,v_2)$ is a basis of $\ker(F-I)^2$. Lastly
$(v_3,\dots v_{2d})$ is a basis of $V_z:=\underset{\la\neq 1}{\oplus} E_\la$.\\
Let us set $w:=\frac{i}{2}(F+I)\grad$. We have $Dv_1=-w$ and, using lemma \ref{combescuretechnic}:
\beq\label{cat}
((F-I) +K^{-1}\Pi)v_2=(F-I)v_2.
\eeq
\beq\label{woo}
\frac{1}{\aaa}\jgrad [i^t\grad +^tX_1(F-I)]v_2=\frac{1}{\aaa}(i<\grad,v_2> + <X_1,(F-I)v_2>)\jgrad.
\eeq
Using the fact that $(F-I)v_2\in E_1$, one easily gets that there exists
$\la_1\in\rr$ such that $(F-I)v_2=\la_1\jgrad$. Thus $<X_1,(F-I)v_2>=\la_1 \aaa.$
We obtain, using (\ref{cat}) and (\ref{woo}):
\beq\label{2emecol}
D v_2=-\frac{i}{\aaa}<\grad,v_2>\jgrad.
\eeq
Note that $(F-I)V_z\subset V_z$. Moreover $K^{-1}\Pi$ is of rank $1$.
Hence, since its image is equal to $K^{-1}\Pi v_1=-w\neq 0$, we can neglect it on others columns than the first
column. The same idea holds for $\frac{1}{\aaa}\jgrad [i^t\grad +^tX_1(F-I)]$, which we neglect in other
columns than the second one (since $\frac{1}{\aaa}\jgrad [i^t\grad +^tX_1(F-I)]v_2\neq 0$). Therefore:
$$\det(D)=\det\left(
\barr{cc|c}
-w_1    & -\frac{i}{\aaa}<\grad,v_2> & 0\\
-w_2    & 0                          & \\\hline
-w_3    & 0                          & \\
\vdots  & \vdots                     & (F-I)_{|_{V_z}} \\
-w_{2d} & 0                          &
\earr\right)$$
where $(w_1,\dots,w_{2d})$ are coordinates of $w$ in basis $\beta_0$.\\
Hence $\det(D)=-\frac{i}{\aaa}w_2<\grad,v_2> \det((F-I)_{|_{V_z}})$.\\
We write
$$w=\frac{i}{2}(F+I)\grad=w_1 \jgrad +w_2 v_2 +v$$
where $v\in V_z$, then we take the scalar product with $\grad$. Since $E_1=(JV_z)^{\perp}$, we have
$<v,\grad>=0$. and $i|\grad|^2=w_2 <v_2;\grad>.$ Thus we get:
$$\det(D)=\frac{1}{\aaa}|\grad|^2 \det((F-I)_{|_{V_z}}).$$
Therefore, according to (\ref{densite2})
\beq\label{densiteND}
d_g(t,z)^{-2}=(-1)^d |\grad|^2 \det((F-I)_{|_{V_z}})\det(\gmu).
\eeq
Since $\det(\gmu)=\pm 1$, there exists $k\in \zz$, depending on $g$, such that: 
$$d_g(t,z)=\frac{e^{ik\frac{\pi}{2} }}{|\grad||\det((F-I)_{|_{V_z}})|^{\ud}}.$$
Moreover, $d_g$ being continuous, $k$ doesn't depend on $z\in\gmm$.
Thus by Theorem \ref{flotpropre}, we have, if $\mathcal{L}_{E,g}:=\{t\in\rr : \exists z\in\nrj\; : \;
\mg\Phi_t(z)=z\}$:
$$I_g(h)=\sum_{t_0\in\mathcal{L}_{E,g}\cap supp(\hat{f})}\sum_{\gmm\in \Gamma_{E,g,t_0}}e^{\frac{i}{h}S_{\gmm}(t_0)}
\frac{\psi(E)\hat{f}(t_0) e^{ik\frac{\pi}{2}}}{2\pi |\det((F-I)_{|_{V_z}})|^{\ud}}
\int_{\gmm}\frac{d\gmm}{|\nabla H|}+O(h).$$
Moreover, if $z\in\gmm$, then:
$$\int_{\gmm}\frac{d\gmm}{|\nabla H|}=\int_0^{T_{\gmm}^*}|J\nabla H(\phi_t(z))|
\frac{dt}{|\nabla H(\phi_t(z))|}=T_{\gmm}^*.$$
$\quad$\\
Lastly, we sum on $g\in G$ to get the expansion of $\mathcal{G}_{\chi}(h)$. This ends the proof of
Theorem \ref{Gutz_osc}.\qed
\section{Appendix : Coherent states}
We recall some basic things on coherent states on $\rdd$ in Schr\"odinger representation. We mainly follow
the presentation of M.Combescure and D.Robert (cf \cite{CR}, \cite{Ro1}).
\subsection{Notations}
The $h$-scaling unitary operator $\Lambda_h:\lde\to \lde$ is defined by:
$$\Lambda_h\psi (x)=\frac{1}{h^{\frac{d}{4}}}\psi\left(\frac{x}{h^{\frac{1}{2}}}\right).$$
The phase translation unitary operator associated to $\aaa=(q,p)=\in\rd\times\rd$ is given by:
$\mathcal{T}_h(\alpha):=\exp [\frac{i}{h}(px-q.hD_x)]$. We classically have
$\trp^*=\trp^{-1}=\mathcal{T}_h(-\alpha)$ and:
\beq\label{etatcoherentboy}
\trp f(x)=\exp\left( i\frac{p}{h}(x-\frac{q}{2})\right).f(x-q).
\eeq
The ground state of the harmonic oscillator $-\Delta +|x|^2$ is given by
$\tilde{\psi_0}(x):=\frac{1}{\pi^{\frac{d}{4}}}\exp (-\frac{|x|^2}{2}).$\\
We set:
\beq
\psi_0(x):=\Lambda_h\tilde{\psi_0}(x)=\frac{1}{(h\pi)^{\frac{d}{4}}}\exp (-\frac{|x|^2}{2h}).
\eeq
Then the coherent state associated to $\aaa\in\rdd$ is given by $\fbox{$\varphi_{\aaa}$}:=\trp\psi_0.$
By (\ref{etatcoherentboy}), we have:
\beq
\varphi_{\aaa}(x)=\frac{1}{(h\pi)^{\frac{d}{4}}}\exp \left(i\frac{p}{h}(x-\frac{q}{2})\right).\exp
\left(-\frac{|x-q|^2}{2h}\right).
\eeq
and we get easily from (\ref{etatcoherentboy}) the following formulae:
$$\chg^*\trp\chg=\mathcal{T}_1(\frac{\aaa}{\sqrt{h}}) \mbox{ and }\chg^*\op(a)\chg= Op^w_1(a_h),
\mbox{ where } \; a_h(z):=a(\sqrt{h}z).$$
$$\trp \mathcal{T}_h(\beta)=e^{\frac{i}{2h}<J\aaa;\beta>}\mathcal{T}_h(\aaa+\beta)\;
\mbox{ and }\;\trp^* \op(a) \mathcal{T}_h(\alpha)=\op[a(\aaa +.)].$$
\subsection{A trace formula} $\quad$\\
If $A\in\mathcal{L}(\lde)$ is trace class, then $\gra|<A\eco;\eco>_{\lde}|d\aaa<+\infty$, and we have:
\beq\label{trace_ec}
\mbox{Tr}(A)=(2\pi h)^{-d}\gra <A\eco;\eco>_{\lde}d\aaa.
\eeq
For a proof, see for example \cite{these}.
\subsection{Propagation of coherent states}
For $z=(q,p)\in\rd\times\rd$, let $z_t=(q_t,p_t):=\Phi_t(z)$ be the solution of
the Hamiltonian system (\ref{hamsyst}) with initial condition $z$. We introduce the notations:
\beq\label{actionman}
S(t,z):=\int_0^t (p_s.\dot{q}_s -H(z_s) )ds
\eeq
\beq
\delta(t,z):=S(t,z)-\frac{q_t p_t-qp}{2}.
\eeq
where $F_\aaa(t)=\partial_z \Phi_t(\aaa)\in Sp\,(d,\rr)$. We set:
\beq\label{monodr}
F_{\aaa}(t)=\left(
\begin{array}{cc}
A & B \\
C & D \\
\end{array}
\right)
\qquad\mbox{ where }\qquad A, B, C, D\in M_d(\rr).
\eeq
\begin{theo}\label{propag_ec}
\underline{Semi-classical propagation of coherent states} (Combescure-Robert) \cite{CR}, \cite{Ro1} :\\
Let $T>0$. Let $H:\rdd\to\rr$ be a smooth Hamiltonian satisfying, for all $\aaa\in\nn^{2d}$ :
\beq\label{hypmin}
|\partial^{\aaa}H(z)|\leq C_{\aaa}<x>^{m_{\aaa}}, \quad \mbox{ where } \; m_{\aaa}>0,\; C_\aaa>0.
\eeq
Let $\aaa\in\rdd$ be such that the solution with initial condition $\aaa$ of the system $\dot{z_t}=J\nabla{H}(z_t)$
is defined for $t\in ]-T,T[$. We denote by $U_h(t):=e^{-i\frac{t}{h}\hq}$
the quantum propagator.\

Then, $\forall M\in\nn, \; \exists C_{M,T}(\aaa)>0$, independant of $h$ and of $t\in[-T,T]$ such that:
$$\norm{U_h(t)\eco-e^{i\frac{\delta(t,\aaa)}{h}} 
\trpt\chg \left[\sum_{j=0}^M h^{\frac{j}{2}} b_j(t,\aaa)(x).e^{\frac{i}{2}<M_0x,x>}\right] }_{\lde}
\leq C_{M,T}(\aaa).\, h^{\frac{M+1}{2}}.$$
where $M_0:=(C+iD)(A+iB)^{-1}$, for all $t\in]-T,T[$, $b_j(t,\aaa):\rd\to\cc$ is a polynomial
independant of $h$, with degree lower than $3j$, with same parity as $j$, and smoothly dependant of
$(t,\aaa)$. In particular, $b_0(t,\aaa)(x)=\pi^{-\frac{d}{4}} (\det (A+iB))_c^{-\frac{1}{2}}$.\

Moreover, if solutions of the Hamiltonian classical system are defined on $[-T,T]$ for initial conditions
$\aaa$ in a compact $K$, then  $\aaa\mapsto C_{M,T}(\aaa)$ is upper bounded on $K$
by $\tilde{C}_{M,T,K}$ independant of $\aaa\in K$.
\end{theo}

\end{document}